\documentclass[prl,aps,superscriptaddress,twocolumn,floats]{revtex4}
\usepackage{epsfig}

\def\ee{\end{equation}}
\def\bea{\begin{eqnarray}}
\def\eea{\end{eqnarray}}

\def\>{\rangle}
\def\<{\langle}

\def\<{\langle}
\def\>{\rangle}

\begin{document}


\title{Kondo Effect in Electromigrated Gold Break Junctions}

\author{A. A. Houck}

\affiliation{Center for Bits and Atoms and Department of Physics,
             MIT, Cambridge, MA 02139}

\affiliation{Department of Physics,
             Harvard University, Cambridge, MA 02138}

\author{J. Labaziewicz}
\affiliation{Center for Bits and Atoms and Department of Physics,
             MIT, Cambridge, MA 02139}

\author{E. K. Chan}
\affiliation{Center for Bits and Atoms and Department of Physics,
             MIT, Cambridge, MA 02139}

\author{J. A. Folk}
\affiliation{Center for Bits and Atoms and Department of Physics,
             MIT, Cambridge, MA 02139}

\author{I. L. Chuang}
\affiliation{Center for Bits and Atoms and Department of Physics,
             MIT, Cambridge, MA 02139}


\date{\today}

\begin{abstract}
We present gate-dependent transport measurements of Kondo impurities
in bare gold break junctions, generated with high yield using an
electromigration process that is actively controlled. Thirty percent
of measured devices show zero-bias conductance peaks. Temperature
dependence suggests Kondo temperatures $\sim 7K$.  The peak
splitting in magnetic field is consistent with theoretical
predictions for g=2, though in many devices the splitting is offset
from $2g\mu_B B$ by a fixed energy.  The Kondo resonances observed
here may be due to atomic-scale metallic grains formed during
electromigration.
\end{abstract}

\maketitle



The physics of many-particle entangled spin states has attracted
recent interest in a variety of systems.  The correlation that may
form at low temperature between a localized electron spin and a
surrounding Fermi sea is one of the most carefully studied of these
states.  This correlation gives rise to the Kondo effect, originally
observed in the 1930's as an enhanced low-temperature resistance in
magnetically-doped metals\cite{meissner_ap30}. Several decades later
a new manifestation of the same physics was discovered in systems
where a charge trap is positioned between two electrical
leads\cite{appelbaum_pr67, shen_pr68, bermon_prb78, gregory_prl92,
ralph_prl94}.

During transport, electrons tunnel from one lead onto the trap and
off to the other lead.  In general, transport is allowed only when
two charge states are degenerate, an effect known as Coulomb
blockade. If the trap binds an unpaired electron, high-order
tunneling processes may give rise to an enhanced conductance at
zero-bias in spite of Coulomb blockade. Metal tunnel junctions
containing paramagnetic impurities were the first systems to show
the narrow zero-bias peak characteristic of this
effect\cite{appelbaum_pr67, shen_pr68, bermon_prb78, gregory_prl92,
ralph_prl94}. The peak splits in magnetic field, and is suppressed
above the Kondo temperature, $T_K$.

The Kondo temperature corresponds to the energy of formation of the
many-body spin state; it depends exponentially on the binding energy
of the trap, $\epsilon_0$, and the coupling to the leads, $\Gamma$:
$T_K \propto e^{-\pi\mid\epsilon_0\mid/2 \Gamma}$\cite{ralph_prl94}.
An electrostatic gate can be used to control $T_K$ by tuning
$\epsilon_0$, and can even turn off the Kondo resonance by changing
the number of bound electrons. A gateable Kondo effect was first
observed in lithographically-defined quantum dots\cite{dgg_nature98,
cronenwett_science98}, where the primary difficulty was achieving a
Kondo temperature high enough to be experimentally accessible.

Larger Kondo temperatures can be found in much smaller systems
consisting of an atomic-scale charge trap between two metallic
leads. However, fabrication of such leads in a gateable geometry is
a formidable lithographic challenge. One successful approach is to
break a single wire into an electrode pair using
electromigration\cite{park_apl99}, the process by which an electric
current causes the atoms in a conductor to move\cite{ho_rep89}. Both
Coulomb blockade and the Kondo effect have been observed in
electromigrated gold junctions with organic molecules deposited on
the surface as charge traps\cite{park_nature02, liang_nature02,
natelson_nano04}.

In this Letter we show that localized states giving rise to the
Kondo effect may be found even in the gated gold junction itself.
Active control over the electromigration process gives a high yield
(up to $30\%$) of devices showing zero-bias peaks that are
suppressed at finite temperature and split with magnetic field. For
most devices, $T_K$ fell in the range $5-10K$.  Above $g\mu_B B\sim
0.5 k_B T_K$ the splitting was linear in field with slope g=2, but
the absolute magnitude of the splitting varied from device to device
for reasons that are not yet understood.  We discuss possible
sources of the Kondo resonance in bare junctions, focusing on
atomic-scale metallic grains.  Grains consisting of $18-22$ atoms,
formed in electromigrated junctions, have been shown to possess
novel optical properties\cite{gonzalez_prl04}; the transport
properties presented here offer a further glimpse of the rich
physics in this system.

\begin{figure}[htbp]
\begin{center}
  \epsfig{file=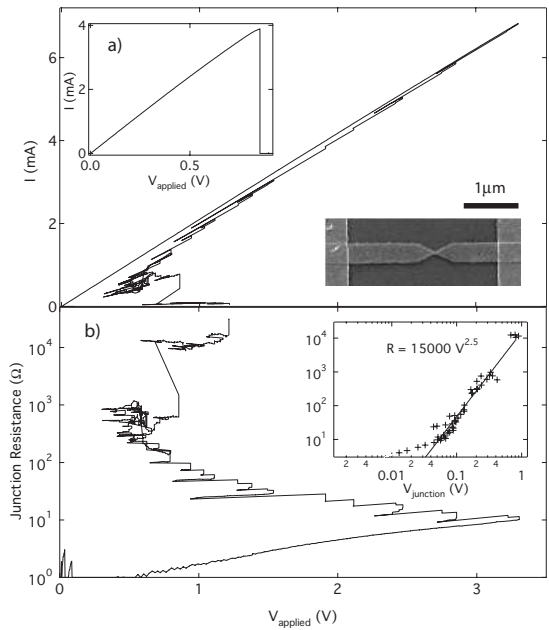, width=0.4\textwidth}
\end{center}
\caption{Two views of controlled electromigration.  (a)
Current-voltage (IV) curve for a characteristic break. Insets show
scanning electron micrograph of device, and IV curve of uncontrolled
electromigration for comparison. (b) Same data, after subtracting
lead resistance.  Voltage dropped in leads is also subtracted in
inset.  Two distinct phases of breaking can be seen: the initial
voltage ramp in which no breaking occurs (A to B) and controlled
electromigration (B to C). The entire process takes $\sim10 min$ for
each wire.  Solid line in inset to (b) shows fit of B-C section to
power law; extracted exponents for all devices suggest $R\propto
V^{2\pm0.5}$.} \label{fig:break}
\end{figure}

The devices (Fig.\,1a inset) are patterned on a GaAs substrate using
a combination of optical and electron beam lithography and a
resist/liftoff process.  A $12nm$ thick aluminum back gate is
deposited by evaporation with the substrate held at $77K$. The
surface is oxidized for ten minutes in ozone under a UV lamp. Gold
wires ($15nm$ thick) are deposited on the gate at room temperature;
each tapers to a $70nm$ constriction.  A $50nm$ gold layer connects
the narrow wires to optical lithography.  The pressure during
evaporation was $\sim2\times10^{-7}mbar$.  The devices are cleaned
with acetone and methanol, followed by oxygen plasma to remove any
residual organic materials. Electromigration is performed in a
$^3He$ refrigerator containing $1mbar$ of $^3He$ exchange gas. The
sample is then cooled immediately and measured.

An active control algorithm differentiates the electromigration
performed here from previous work. When junctions are created
without using feedback, the yield showing even weak tunneling is low
($5-10\%$) and there is little control over final resistance. For
better control we monitor the resistance of the entire
electromigration circuit and adjust the voltage to maintain a
constant break rate, measured as $(1/R) (\partial R/\partial t)$,
where $R$ is the resistance of the circuit after subtracting a lead
resistance. Break rate is extracted from $20$ resistance
measurements at fixed voltage, taken over $200ms$.

\begin{figure}[floatfix]
\begin{center}
\epsfig{file=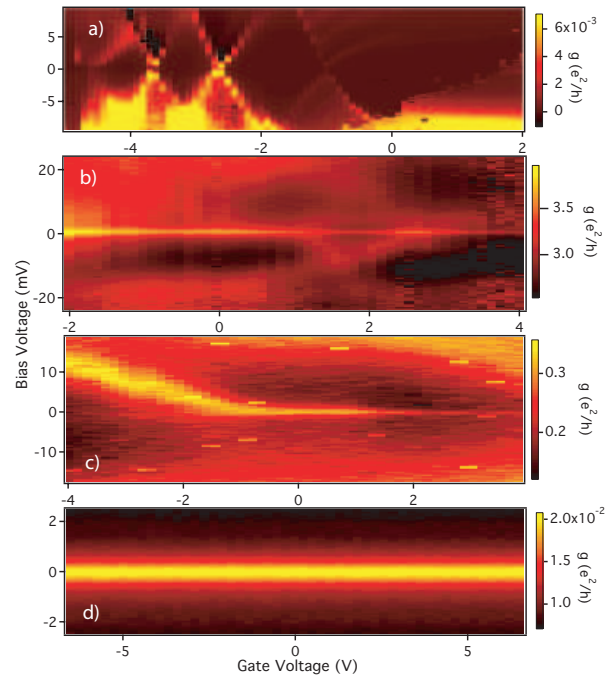,width=0.44\textwidth, bb = 15 0 421 457}
\end{center}
\caption{Differential conductance maps for four devices. (a) Coulomb
blockade. (b) Superposition of a broad Coulomb blockade diamond
(centered around gate voltage $1.8V$) and the Kondo effect.  (c)
Transition from Kondo resonance to Coulomb diamond. The absence of
conductance features at negative bias may be due to an asymmetric
lead coupling. (d) Gate-independent Kondo effect. (c) is numerical
derivative of DC measurements; others measured with $50\mu V$ AC
bias at $47 Hz$. } \label{fig:twods}
\end{figure}

Figure 1 shows the two stages of a typical break.  In the first (A
to B), the break rate is negligible as the voltage is ramped up.
Once the break rate exceeds a target ($0.01-0.1s^{-1}$), the voltage
is actively adjusted to maintain that rate (B to C). When the
resistance exceeds $50k\Omega$, the voltage is reduced to zero at
$1V/s$.  There is remarkable consistency from device to device;
extracted power laws indicate that constant power is dissipated in
the break region for the B-C stage (see Fig.\,1b inset).

We present results for electromigration of $167$ wires, of which
$118$ broke to a room-temperature resistance between $10k\Omega$ and
$5M\Omega$.  We cooled $72$ of these to $250mK$ and measured
differential conductance as a function of gate voltage and DC
bias\cite{why72}. No conductance, or only weakly nonlinear
conductance, was visible in $31$ devices. Three showed a
gate-independent conductance dip at zero bias suggesting transport
through metallic grains in good contact with the
leads\cite{anaya_jap03}.

The remaining 38 showed transport through charge traps: Coulomb
blockade or a Kondo resonance (see Fig.\,2).  The binding energy of
these traps is primarily determined by the energy, $E_C$, needed to
add or remove an electron from a trap. No device that showed a Kondo
resonance also showed two charge degeneracy points, necessary for
accurate measurement of $E_C$; in many the conductance was
featureless up to hundreds of $mV$.  However, multiple Coulomb
diamonds were occasionally observed in devices without Kondo
resonances, giving values as low as $E_C\sim5mV$ in these
(e.g.\,Fig.\,2a).

\begin{figure}[floatfix]
\begin{center}
  \includegraphics[width=0.44\textwidth]{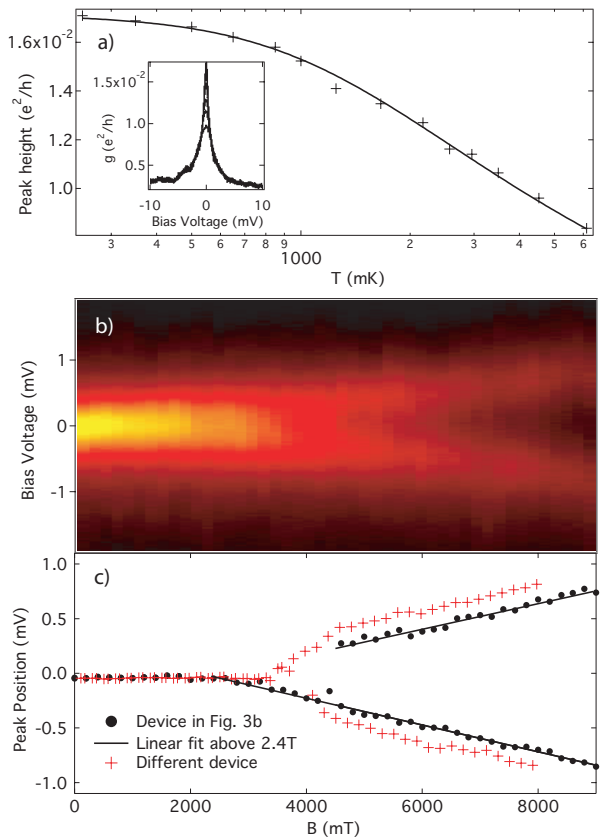}
\end{center}
\caption{Signatures of the Kondo effect in a zero-bias conductance
peak. (a) Suppression with temperature.  Peak heights are extracted
from raw data in inset.  Fit to theoretical form (solid curve) gives
$T_K = 6K$ for this device. (b) Splitting in magnetic field. (c)
Extracted peak positions from the data in (b) (black circles) and
from a different junction with the same Kondo temperature (red
crosses). Solid black lines are linear fits to peak position.
Splitting, defined as distance between peak positions, is linear for
both devices (slope $2 g \mu_B$ with $g = 2.1\pm 0.05$) but
extrapolates to zero at different fields.  } \label{fig:kondo}
\end{figure}

A zero-bias conductance peak was observed in $30$ devices. One
signature of the Kondo effect is a suppression of the peak
conductance with temperature: $G(T) \simeq
G(0)/[1+(2^{1/s}-1)(T/T_K)^2]^s$, where s = 0.22 for spin-1/2
impurities\cite{costi_94, dgg_prl98}. We measured temperature
dependence in $9$ peaks; all showed a good fit to this functional
form, see e.g.\,Fig.\,3a. The possibility of conductance pathways in
parallel with the Kondo resonance (Fig.\,2b) prevents reliable
determination of $T_K$ from temperature dependence alone, so the
width of the zero bias peak was used as confirmation. In most
devices the peak width (defined as full width half maximum) was $2
k_B T_K$\cite{costi_condmat}. We consider $T_K$ to be
well-determined only in these devices.   For most, $T_K\sim7K$; none
showed $T_K<4K$, and only one $T_K>15K$.

Gate dependence of the Kondo resonance was evident in $26$ of the
$30$ devices, but charge degeneracy points, identifiable by a
Coulomb diamond and representing a change in the number of bound
electrons, were accessible in only three. In one, the zero-bias peak
appears on only one side of the degeneracy point in gate voltage, as
expected (see Fig.\,2c). In the other two devices, however, the
zero-bias peaks continue through the edge of the diamonds
(Fig.\,2b), leading us to suspect two parallel but independent
conduction pathways.

Another identifying characteristic of the Kondo effect is the
splitting of the zero-bias peak by $2 g\mu_B B$ in magnetic
field\cite{appelbaum_pr67}.  Field dependence was measured in $12$
devices, and this signature was observed in all\cite{explainfields}.
The splitting was visible only above a device-dependent critical
field, $B_c$.  Where $T_K$ was well-determined, $g \mu B_c \sim 0.5
k_B T_K$, consistent with theoretical predictions\cite{Costi_prl00}.

Above $B_c$ the splitting was linear; typical data are shown in
Fig.\,3. In 8 devices the slope of the splitting was consistent with
$g=2.1\pm0.1$.  The other four showed g-factors up to 4.5; these
anomalously large values are not yet understood. In contrast,
theoretical work predicts a \emph{suppression} of the g factor, to
between 1.3 and 1.7, for $g\mu_B B$ up to several times $k_B
T_K$\cite{moore_prl00, costi_condmat}.

The measurements in this paper cover the regime $g\mu_B B\sim k_B
T_K$, where exact theoretical predictions for conductance have not
been worked out.  We find for the splitting of the zero bias peak a
surprisingly simple functional form: $2g\mu_B (B-B_{ext})$, where
$B_{ext}$ is device dependent. Even devices with the same $g$ factor
and Kondo temperature show different values for $B_{ext}$ (see
e.\,g.\,Fig.\,3c). A negative value, $B_{ext}=-2.5T$, was observed
in one device, reminiscent of a recent observation in GaAs quantum
dots\cite{kastner_condmat}.

A structure that consists simply of two gold electrodes above an
electrostatic gate would not be expected to show either Coulomb
blockade or the Kondo effect. The system must also contain one or
more charge traps.  While there are many candidates for charge traps
in an electromigrated junction, the trends observed here may
identify this additional component.  Changes in the details of the
electromigration procedure dramatically affect the yield of devices
with zero-bias peaks, suggesting a strong dependence on the
morphology of the gold. However, the range of $T_K$ is surprisingly
narrow given the inherent randomness of electromigration and the
sensitivity of $T_K$ to microscopic device
parameters\cite{ralph_prl94}.

Defects in the gate oxide are an unlikely candidate for the Kondo
resonance here, though they have been shown to give rise to the
Kondo effect in other systems\cite{appelbaum_pr67, shen_pr68,
bermon_prb78, gregory_prl92, ralph_prl94}.  Some devices were gate
independent, suggesting that the Kondo impurities can be in a
location screened by the gold leads. In addition, the location of
defect sites relative to the leads would vary from device to device,
giving a wide range of $T_K$.

Atomic or molecular adsorbates on the surface of the
gold\cite{yu_nano04} are a more plausible, but still unlikely,
source of Kondo impurities. The coupling between an adsorbed
molecule and a lead would depend only weakly on junction geometry,
consistent with our results. Oxygen has been cited as a source of
Kondo impurities in metallic tunnel junctions\cite{bermon_ssc}. This
possibility, however, does not explain the observed sensitivity to
electromigration parameters.  In addition, we performed the same
process on a small number of devices in ultra-high vacuum (with the
walls of the vacuum chamber held at $4K$) and found similar device
characteristics.

Isolated metal grains created during electromigration are the final
possible Kondo impurity that we consider.  Coulomb blockade in
grains of size $\gtrsim 2nm$ has been studied
extensively\cite{ralph_prl95}, and observed occasionally in bare
electromigrated junctions\cite{natelson_nano04}.  However, even over
a wide range of lead-grain couplings\cite{anaya_cm}, no observation
of the Kondo effect has been reported.

In principle a metal grain, if well-connected to the leads, could
show a Kondo temperature as high as those reported here.  However,
even in the limit of nearly perfect lead-grain contact, one expects
$T_K\propto\Delta^{3/2}/E_C^{1/2}$ for a grain with quantum level
spacing $\Delta$ \cite{glazman_prl99}. A rough geometric argument
for a grain of radius $r$ then gives $T_K\propto r^{-7/2}$
\cite{ralph_prl95}; even a modest range of sizes would result in a
large range of $T_K$, in contrast with our observations.  In
addition, gold grains larger than a few $nm$ would exhibit a $g$
factor much less than 2 due to spin-orbit
interactions\cite{petta_prl01}.

A more likely candidate for the Kondo impurity in these devices is
an atomic-scale gold nanocluster of the type reported in
Ref.\,\cite{gonzalez_prl04}.  Those nanoclusters, formed during
electromigration, contained $18-22$ atoms\cite{gonzalez_prl04,
lee_pnas02}.  This narrow range of sizes could explain the Kondo
temperatures observed here, and the formation of these structures
would depend strongly on the electromigration procedure.
Atomic-scale nanoclusters are in many ways more similar to molecules
than to larger gold grains, and could display many of the physical
effects expected in single-molecule transport. The electronic
measurements of many-body spin correlations presented here, together
with the recent demonstration of single photon generation in these
nanoclusters\cite{gonzalez_prl04}, show the promise that this system
may hold for coherent electronics in the solid state.

Electromigrated gold junctions have been used by many groups to
measure transport through single organic molecules.  Control
experiments on bare gold wires were used to distinguish between
transport features due to the molecules and those native to the
electrode system.  The sensitivity of our results to the details of
electromigration demonstrates the difficulty in interpreting these
control experiments, as simply the presence of molecules deposited
on the surface may change the morphology of the gold. This
sensitivity can also be an advantage: we find that the yield of
Kondo resonances in bare junctions can be dramatically reduced by
briefly applying a large voltage ($>1V$) across the junctions at
room temperature. This large voltage is a natural result of more
common electromigration procedures.

In summary, we have observed the Kondo effect in bare
electromigrated gold junctions.  The mere presence of Kondo
resonances in this system is surprising and warrants further study.
Careful analysis of the splitting at magnetic fields comparable to
the Kondo temperature reveals a simple but unexpected functional
form. Experiments are underway to better understand this behavior.
The electromigration procedure outlined in this paper can also be
extended to superconducting materials, and offers an experimental
platform to test emerging theoretical models of the interplay
between the Kondo effect and superconductivity.

{\small {\bf Acknowledgements:} This work is funded by an HP-MIT
  alliance through the Quantum Science Research Group, AFOSR MURI
  Award no. F49620-03-1-0420, and the NSF Center for Bits and Atoms.
  The authors thank T. Costi, D. Davidovic, G. Fiete, L. Glazman, D. Goldhaber-Gordon,
  H. Heersche, W. Hoftstetter, M. Kastner, C. Marcus, J. Moore, D. Natelson, A. Pasupathy, D. Stewart and S. Williams for valuable
  discussions and advice.  AAH acknowledges support from
  the Hertz Foundation.  JAF acknowledges support through a Pappalardo
  Postdoctoral Fellowship.}

\bibliographystyle{apsrev}
\bibliography{aukondo16a}

\begin{thebibliography}{30}
\bibitem[{why()}]{why72}
\bibinfo{note}{Devices not cooled were used to test breaking procedure.}
\bibitem[{exp()}]{explainfields}
\bibinfo{note}{Field dependence in a total of 15 devices was measured, but two
  were too noisy to extract a reliable splitting and one had too high a Kondo
  temperature: $T_K = 34K$.}
\end{thebibliography}

\newpage

\end{document}